\documentclass[a4paper,11pt]{article}
\pdfoutput=1 

\usepackage{jheppub} 

\usepackage[T1]{fontenc} 

\title{\boldmath Joule-Thomson Expansion of Kerr$-$AdS Black Holes}

\author[a,1]{\"{O}zg\"{u}r \"{O}kc\"{u}\note{Corresponding author.}}
\author[a]{ and Ekrem Ayd{\i}ner}

\affiliation[a]{Department of Physics, Faculty of Science, Istanbul University\\Vezneciler 34134 Istanbul, Turkey}

\emailAdd{ozgur.okcu@ogr.iu.edu.tr}
\emailAdd{ekrem.aydiner@istanbul.edu.tr}

\abstract{In this paper, we study Joule-Thomson expansion for Kerr$-$AdS black holes in the extended phase space. A Joule-Thomson expansion formula of Kerr$-$AdS black holes is derived. We investigate both isenthalpic and numerical inversion curves in the $T-P$ plane and demonstrate the cooling-heating regions for Kerr$-$AdS black holes. We also calculate the ratio between minimum inversion and critical temperatures for Kerr$-$AdS black holes.}

\begin{document} 
\maketitle
\flushbottom

\section{Introduction}
\label{sec:intro}

Since the first studies of Bekenstein and Hawking \cite{Bek1972,Bek1973,Bard1973,Hawk1974,Bek1974,Hawk1975}, black holes as thermodynamic system have been an interesting research field in theoretical physics. The black hole thermodynamics provides fundamental relations between theories such as classical general relativity, thermodynamics and quantum mechanics. Black holes as thermodynamic system have many exciting similarities with general thermodynamic system. These similarities become more obvious and precise for the black holes in AdS space. The properties of AdS black hole thermodynamics have been studied since the seminal paper of Hawking and Page \cite{Hawk1983}. Furthermore, the charged AdS black holes thermodynamic properties were studied in \cite{Cham1999a,Cham1999b} and it was shown that the charged AdS black holes have a van der Waals like phase transition.

Recently black hole thermodynamics in AdS space has been intensively studied in the extended phase space where the  cosmological constant is considered as the thermodynamic pressure. Extended phase space leads to important results: Smarr relation is satisfied for the first law of the black holes thermodynamics in the presence of variable cosmological constant. It also provides the  definition of the thermodynamic volume which is more sensible than the geometric volume of the black hole. In addition to similar behaviours with conventional thermodynamic systems,  studying the AdS black holes is another important reason for the AdS$/$CFT correspondence \cite{Juan1999}. Considering the cosmological constant as thermodynamic pressure,
\begin{equation}
	\label{pressure}
	P=-\frac{\Lambda}{8\pi}\,,
\end{equation}
and its conjugate quantity as thermodynamic volume
\begin{equation}
	V=\left(\frac{\partial M}{\partial P}\right)_{S,Q,J}
\end{equation}
lead us to investigate thermodynamic properties, rich phase structures and other thermodynamic phenomena for AdS black holes in a similar way to the conventional thermodynamic systems.

Based on this idea, the charged AdS black hole thermodynamic properties and phase transition were studied by Kubiznak and Mann \cite{Kub2012}. It was shown in this study that the charged AdS black hole phase transition has the same characteristic behaviors with van der Waals liquid-gas phase transition. They also computed critical exponents and showed that they coincide with exponents of van der Waals fluids. It was shown in \cite{Kast2009} that cosmological constant as pressure requires considering the black hole mass $M$ as the enthalpy $H$ rather than as internal energy $U$. In recent years, thermodynamic properties and phase transition of AdS black holes  have been widely investigated \cite{Dolan2011,Dolan2012,Mo2013,Wei2016,Guna2012,Spal2013,Bel2013,Cai2013,Zha2013,Ma2014,Belhaj2015,Dutta2013,Li2014,Liang2016,Hendi2013,Hendi2016,Sadeghi2014,Mome2017,Alta2014b,Frass2014,Henn2015,Wei2015,Dolan2011b,Dolan2014,Maj2017,John2014,John2016,John2016b,Bel2015b,Caceres2015,Jaf2017,Setare2015,John2016c,Bham2017,Liu2017,Mo2017,Zhang2016,Dolan2015,Alta2014,Kub2017,Lan2017,Wei2017}.\footnote{See  \cite{Dolan2015,Alta2014,Kub2017} and references therein for various black hole solutions.} The phase transition of AdS black holes in the extended phase space is not restricted to van der Waals type transition, but also reentrant phase transition and triple point for AdS black holes were studied in \cite{Alta2014b,Frass2014,Henn2015,Wei2015}. The compressibility of rotating AdS black holes in four and higher dimensions was  studied in \cite{Dolan2011b,Dolan2014}.  In \cite{Maj2017}, a general method was used for computing the critical exponents for AdS black holes which have van der Waals like phase transition. Furthermore, heat engines behaviours of the AdS black holes have been studied. For example, in \cite{John2014}  two kind of heat engines were proposed by Johnson for charged AdS black holes and heat engines were studied for various black hole solutions in \cite{John2016,John2016b,Bel2015b,Caceres2015,Jaf2017,Setare2015,John2016c,Bham2017,Liu2017,Mo2017,Zhang2016}.\footnote{See \cite{Kub2017} and references therein.} More recently, adiabatic processes \cite{Lan2017} and Rankine cycle \cite{Wei2017} have been studied for the charged AdS black holes. 

In \cite{Okcu2017}, we also studied the well-known Joule-Thomson expansion process for the charged AdS black holes. We obtained inversion temperature to investigate inversion and isenthalpic curves. We also showed heating-cooling regions in $T-P$ plane. However, so far, Joule-Thomson expansion for Kerr-AdS black holes in extended phase space has never been studied. The main purpose of this study is to investigate Joule-Thomson expansion for Kerr-AdS black holes.

The paper is arranged as follows. In Sect. \ref{KBH}, we briefly review some thermodynamic properties of Kerr-AdS black holes which are introduced in \cite{Dolan2012}\footnote{Indeed, Kerr-Newman-AdS black hole thermodynamics functions are introduced in \cite{Dolan2012}. But one can easily obtain Kerr$-$AdS black holes thermodynamic functions, when electric charge $Q$ goes to zero.}. In Sect. \ref{JTE}, we first of all derive a Joule-Thomson expansion formula for Kerr-AdS black hole by using first law and Smarr formula. Then we obtain equation of inversion pressure $P_{i}$ and entropy $S$ to investigate the inversion curves. We also show that the ratio between minimum inversion and critical temperatures for Kerr$-$AdS black holes is the same as the ratio of charged AdS black holes \cite{Okcu2017}. Finally, we discuss our results in Sect. \ref{Con}. (Here we use the units $G_{N}=\hbar=k_{B}=c=1$.)

\section{Kerr$-$AdS Black Holes}
\label{KBH}
In this section, we briefly review Kerr$-$AdS black hole thermodynamic properties in the extended phase space. The line element of Kerr$-$AdS black hole in four dimensional AdS space is given by
\begin{eqnarray}
	\label{kerrMetric}
	ds^{2}=-\frac{\Delta}{\rho^{2}}\left(dt-\frac{a\sin^{2}\theta}{\Xi}d\phi\right)^{2}+\frac{\rho^{2}}{\Delta}dr^{2}+\frac{\rho^{2}}{\Delta_{\theta}}d\theta^{2}+\frac{\Delta_{\theta}\sin^{2}\theta}{\rho^{2}}\left(adt-\frac{r^{2}+a^{2}}{\Xi}d\phi\right)^{2},
\end{eqnarray}
where
\begin{eqnarray}
	\label{kerrParameters}
	&\Delta&=\frac{(r^{2}+a^{2})(l^{2}+r^{2})}{l^{2}}-2mr \,,
	\qquad
	\Delta_{\theta}=1-\frac{a^{2}}{l^{2}}\cos^{2}\theta \,,
	\nonumber\\
	&\rho^{2}&=r^{2}+a^{2}\cos^{2}\theta \,, \qquad
	\Xi=1-\frac{a^{2}}{l^{2}} \,,
\end{eqnarray}
and $l$ represents AdS curvature radius. The metric parameters $m$ and $a$ are related to the black hole mass $M$ and the angular momentum $J$ by
\begin{equation}
	\label{fundamentalParameters}
	M=\frac{m}{\Xi^{2}} \,,
	\qquad
	J=a\frac{m}{\Xi^{2}} \,.
\end{equation}
The mass of Kerr$-$AdS black hole in terms of $S$, $J$ and $P$ \cite{Cald2000,Dolan2012} is given by
\begin{equation}
	\label{mass}
	M=\frac{1}{2}\sqrt{\frac{\left(S+\frac{8PS^{2}}{3}\right)^{2}+4\pi^{2}\left(1+\frac{8PS}{3}\right)J^{2}}{\pi S}}\,.
\end{equation}
The first law and the corresponding Smarr relation of Kerr-AdS black hole are given by
\begin{equation}
	\label{firstLaw}
	dM=TdS+VdP+\Omega dJ \,,
\end{equation}
\begin{equation}
	\label{Smarr}
	\frac{M}{2}=TS-VP+\Omega J \,,
\end{equation}
respectively, and the Smarr relation can be derived by a scaling argument \cite{Kast2009}. From Eq.(\ref{firstLaw}), one can obtain the thermodynamic quantities. The expression for the temperature is 
\begin{equation}
	\label{Tempature}
	T=\left(\frac{\partial M}{\partial S}\right)_{J,P}=\frac{1}{8\pi M}\left[\left(1+\frac{8PS}{3}\right)\left(1+8PS\right)-4\pi^{2}\left(\frac{J}{S}\right)^{2}\right] \,.
\end{equation}
Thermodynamic volume is defined by
\begin{equation}
	\label{Volume}
	V=\left(\frac{\partial M}{\partial P}\right)_{S,J}=\frac{2}{3\pi M}\left[S\left(S+\frac{8PS^{2}}{3}\right)+2\pi^{2}J^{2}\right] \,.
\end{equation}
Finally, we obtain angular velocity as follows:
\begin{equation}
	\label{angVelo}
	\Omega=\left(\frac{\partial M}{\partial J}\right)_{S,P}=\frac{\pi J}{MS}\left(1+\frac{8PS}{3}\right) \,.
\end{equation}

In this section, we obtain some thermodynamic quantities of Kerr$-$AdS black holes. In the next section, we will use these quantities to investigate Joule-Thomson expansion effects for Kerr$-$AdS black holes.

\section{Joule-Thomson Expansion}
\label{JTE}
In this section, we will investigate Joule$-$Thomson expansion for Kerr$-$AdS black holes. The expansion is characterized by temperature change with respect to pressure. Enthalpy remains constant during the expansion process. As we know from \cite{Kast2009}, black hole mass is identified enthalpy in AdS space. Therefore, the black hole mass remains constant during expansion process. Joule$-$Thomson coefficient $\mu$, which  characterizes the expansion, is given by \cite{Maytal1997}
\begin{equation}
	\label{JTC1}
	\mu=\left(\frac{\partial T}{\partial P}\right)_{J,M}\,.
\end{equation}
Cooling-heating regions  can be determined by sign of Eq.(\ref{JTC1}). Change of pressure is negative since the pressure always decreases during the expansion. The temperature may decrease or increase during process. Therefore temperature determines sign of $\mu$. If $\mu$ is positive (negative), cooling (heating) occurs. The inversion curve, which is obtained at $\mu=0$ for infinitesimal pressure drops, characterizes the expansion process and it determines the cooling$-$heating regions in $T-P$ plane \footnote{There are two approaches for the Joule$-$Thomson expansion process. The differential and integral versions correspond to infinitesimal and finite pressure drops, respectively. In this paper, we considered differential version of Joule$-$Thomson expansion for Kerr$-$AdS black holes. See \cite{Maytal1997}. }. 

We begin to derive Joule$-$Thomson expansion coefficient formula for Kerr$-$Ads black holes. First, we differentiates Eq.(\ref{Smarr}) to obtain 
\begin{equation}
	\label{DSr}
	dM=2(TdS+SdT-VdP-PdV+\Omega dJ+Jd\Omega)\,.
\end{equation}
Since $dM$=$dJ$=0, Eqs.(\ref{firstLaw}) and (\ref{DSr}) can be written
\begin{equation}
	\label{JTFL}
	TdS=-VdP\,,
\end{equation}
\begin{equation}
	\label{DSr2}
	TdS+SdT-VdP-PdV+Jd\Omega=0\,,
\end{equation}
respectively. If Eq.(\ref{JTFL}) can be substituted into Eq.(\ref{DSr2}), one can obtain
\begin{equation}
	\label{SfKF}
	-2V+S\left(\frac{\partial T}{\partial P}\right)_{M}-P\left(\frac{\partial V}{\partial P}\right)_{M}+J\left(\frac{\partial\Omega}{\partial P}\right)_{M}=0\,,
\end{equation}
which can be rearranged to give Joule$-$Thomson formula as follows:
\begin{equation}
	\label{jTFBH}
	\mu=\left(\frac{\partial T}{\partial P}\right)_{M}=\frac{1}{S}\left[P\left(\frac{\partial V}{\partial P}\right)_{M}-J\left(\frac{\partial\Omega}{\partial P}\right)_{M}+2V\right]\,.
\end{equation}
Here we obtain Joule$-$Thomson expansion formula in terms of Kerr$-$AdS black hole parameters. At inversion pressure $P_{i}$, $\mu$ equals zero and therefore we obtain $P_{i}$ from Eq.(\ref{jTFBH}),
\begin{equation}
	\label{inversionPressure}
	P_{i}=\left(\frac{\partial P}{\partial V}\right)_{M}\left[J\left(\frac{\partial\Omega}{\partial P}\right)_{M}-2V\right]\,.
\end{equation}
From Eq.(\ref{mass}), we can obtain pressure as function of mass, entropy and angular momentum,
\begin{equation}
	\label{pressure2}
	P=\frac{3}{8}\left[\frac{2\sqrt{\pi}\sqrt{\pi^{3}J^{4}+M^{2}S^{3}}-2\pi^{2}J^{2}}{S^{3}}-\frac{1}{S}\right]\,.
\end{equation}
If we combine Eqs.(\ref{Volume}), (\ref{angVelo}) and (\ref{pressure2}) with Eq.(\ref{inversionPressure}), we obtain a relation between inversion pressure and entropy as follows:
\begin{eqnarray}
	\label{relationSP}
	&256P_{i}^{3}S^{7}+256P_{i}^{2}S^{6}+84P_{i}S^{5}+(9-384\pi^{2}J^{2}P_{i}^{2})S^{4}\nonumber\\&-336\pi^{2}J^{2}P_{i}S^{3}-72\pi^{2}J^{2}S^{2}-72\pi^{4}J^{4}=0\,.
\end{eqnarray}
The last equation is useful to determine inversion curves, but first we will investigate minimum inversion temperature. Eq.(\ref{relationSP}) can be given for $P_{i}=0$
\begin{equation}
	\label{minimumInversionStep}
	S^{4}-8\pi^{2}J^{2}S^{2}-8\pi^{4}J^{4}=0\,,
\end{equation}
and we find four roots for this equation. However, one root is physically meaningful. This root is given by
\begin{equation}
	\label{realroot}
	S=\sqrt{2(2+\sqrt{6})}\pi J\,.
\end{equation}
One can substitute Eq.(\ref{realroot}) into Eq.(\ref{Tempature}) and obtain minimum inversion temperature,
\begin{equation}
	\label{Tmin}
	T_{i}^{min}=\frac{\sqrt{3}}{4(916+374\sqrt{6})^{\frac{1}{4}}\pi\sqrt{J}}\,.
\end{equation}
For Kerr$-$AdS black holes, critical temperature $T_{c}$ is given by \cite{Wei2016}
\begin{eqnarray}
	T_{c}&=&\frac{64k_{1}^{2}k_{2}^{4}+32k_{1}k_{2}^{3}+3k_{2}^{2}-12}{4\pi k_{2}\sqrt{k_{2}(8k_{1}k_{2}+3)(8k_{1}k_{2}^{3}+3k_{2}^{2}+12)}}\frac{1}{\sqrt{J}}\nonumber\\&\simeq&\frac{0.041749}{\sqrt{J}}\,,
\end{eqnarray}
where
\begin{eqnarray}
	k_{1}&=&\frac{1}{64\left(103-3\sqrt{87}\right)^{17/3}}\nonumber\\&\times&\bigg(-2^{2/3}\left(225679003807-24183767608\sqrt{87}\right)\nonumber\\&&
	\sqrt[3]{103-3 \sqrt{87}}-17\left(103-3\sqrt{87}\right)^{2/3}\nonumber\\&&
	\left(484826973\sqrt{87}-5116133497\right)\nonumber\\&&-\sqrt[3]{2}\left(68098470527+5855463275
	\sqrt{87}\right)\bigg),\nonumber\\
	k_{2}&=&\frac{2}{3}\left(2+\sqrt[3]{206-6\sqrt{87}}+\sqrt[3]{206+6\sqrt{87}}\right).\nonumber
\end{eqnarray}
The ratio between minimum inversion and critical temperature is given by
\begin{equation}
	\label{ratio}
	\frac{T_{i}^{min}}{T_{c}}\approx0.504622\,,
\end{equation}
\begin{figure}
	\begin{minipage}{\columnwidth}
		\centering
		\includegraphics[width=10cm]{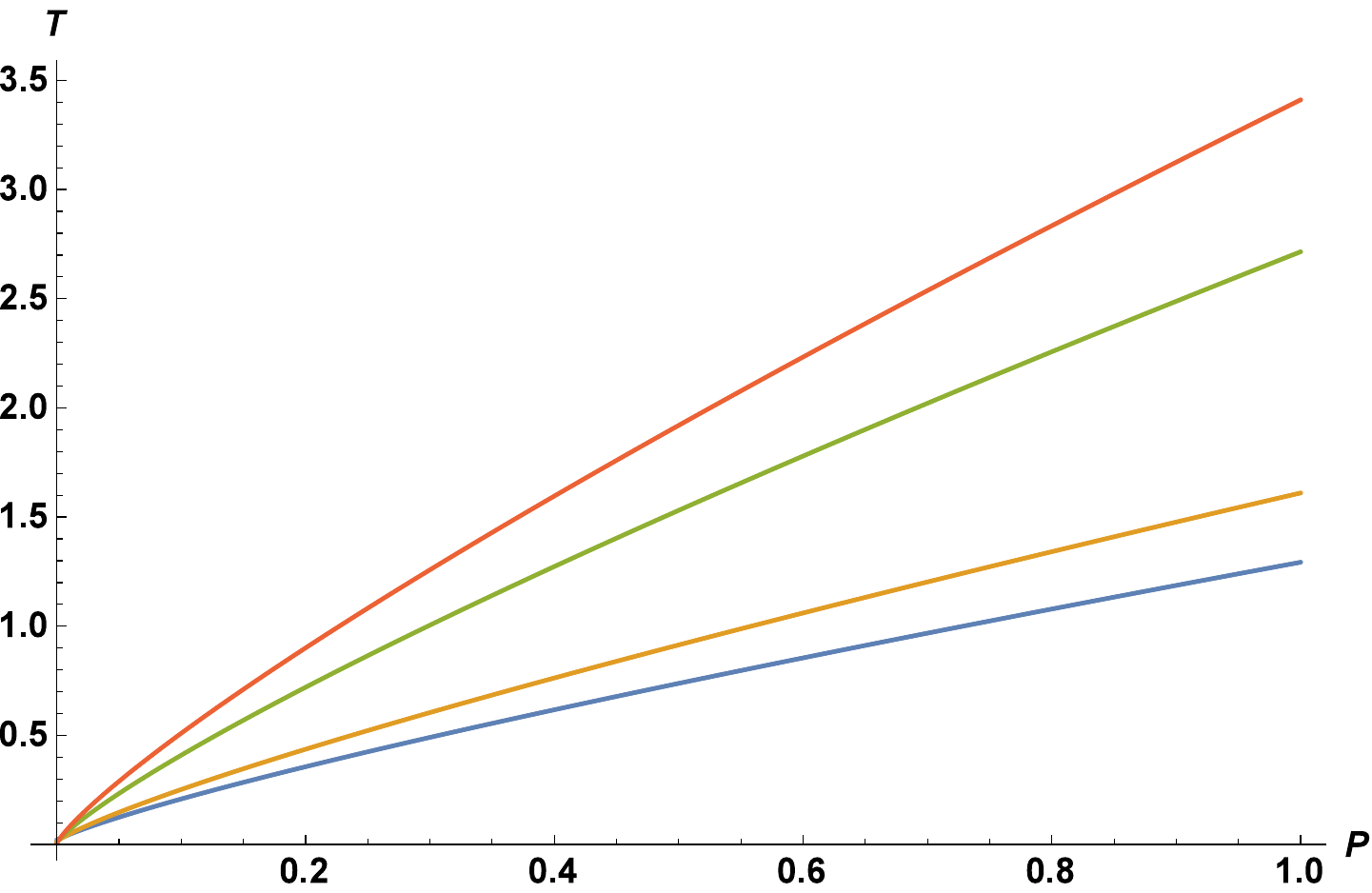}
	\end{minipage}
	\caption{Inversion curves of Kerr$-$Ads black hole in $T-P$ plane. From bottom to top,the curves correspond to $J=1,2,10,20$.}
	\label{iC}
\end{figure}
\begin{figure}[h]
	\centering
	\includegraphics[width=0.47\linewidth]{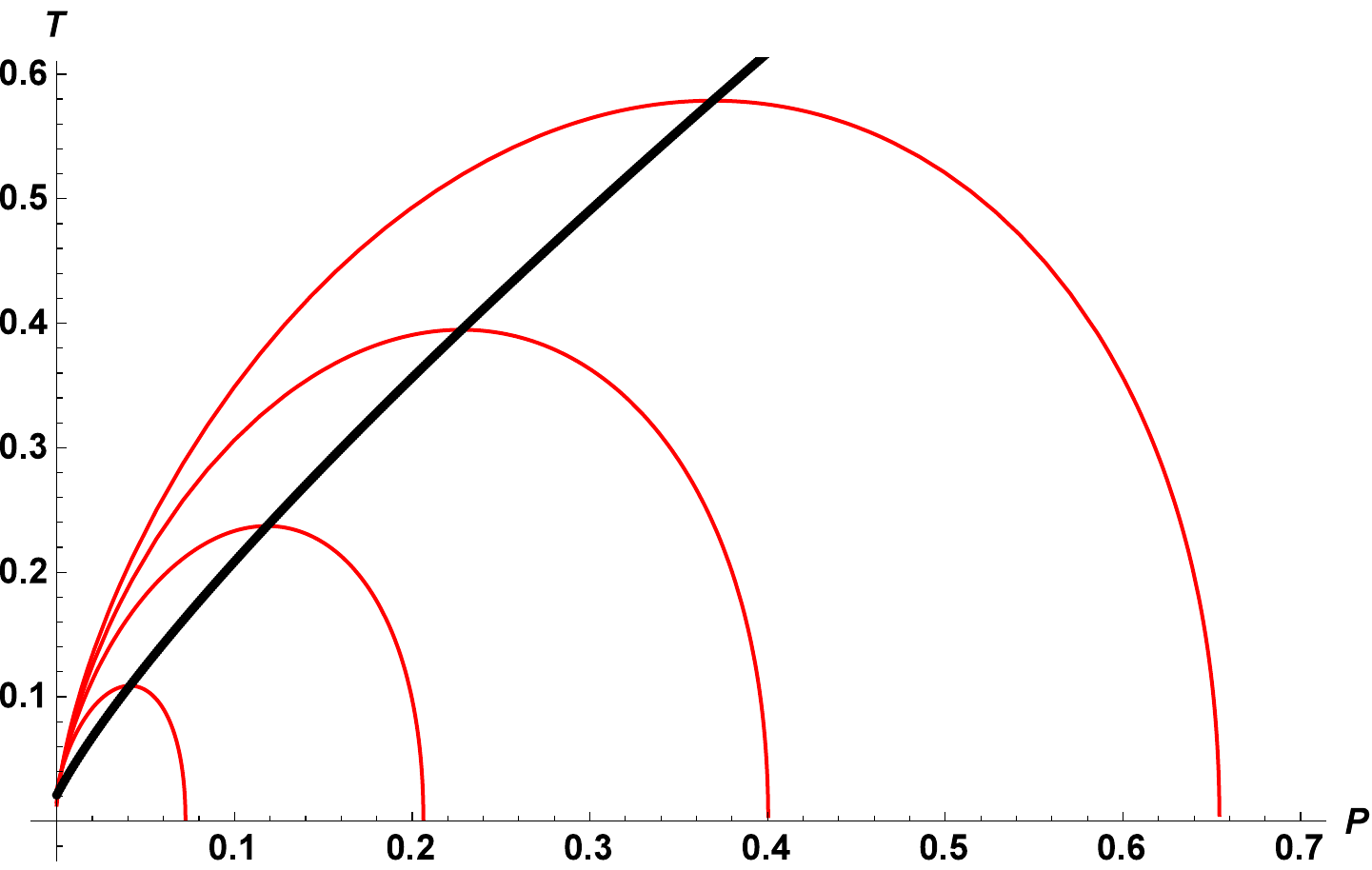}\hfil
	\includegraphics[width=0.47\linewidth]{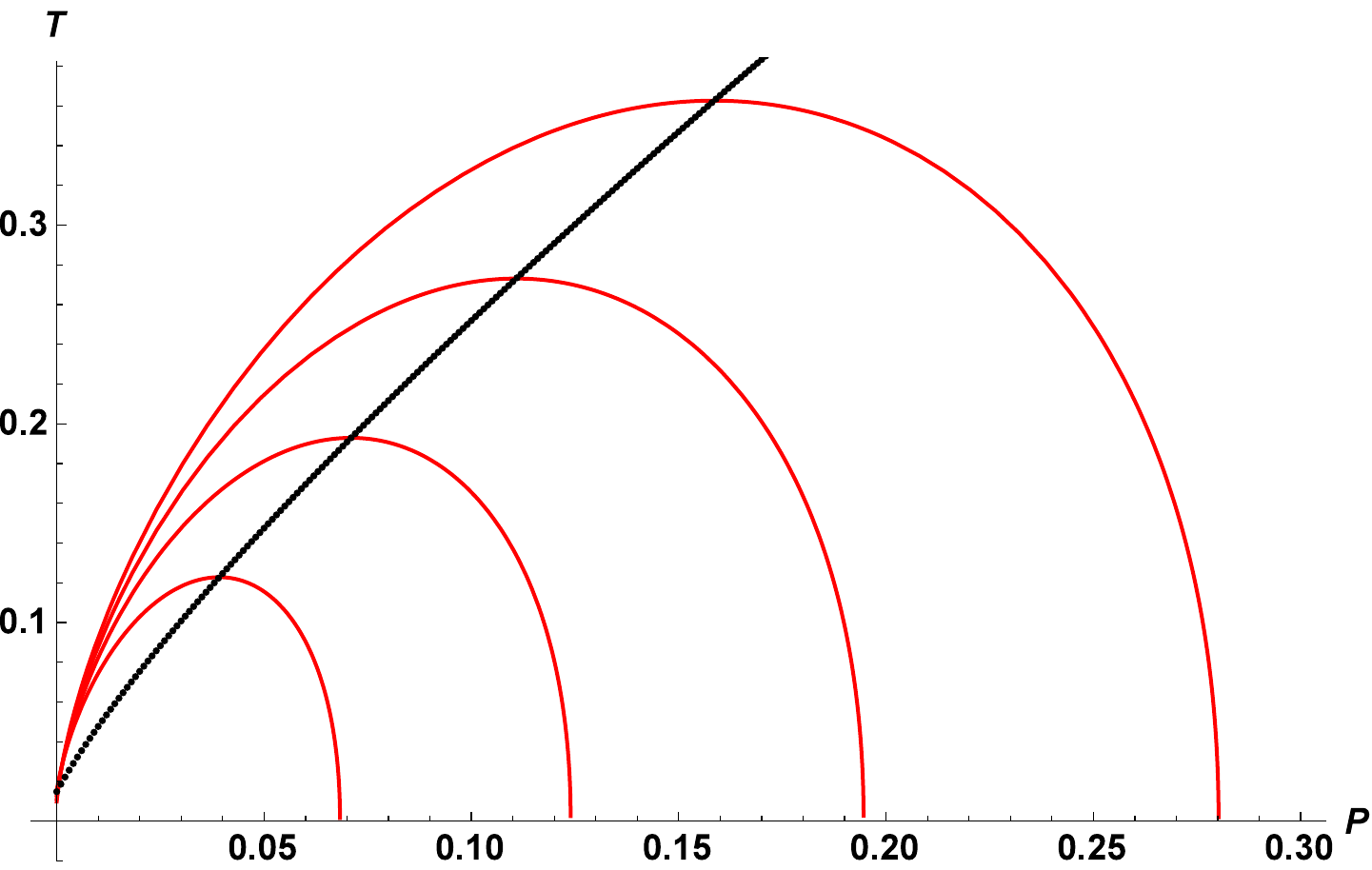}\par\medskip
	\includegraphics[width=0.47\linewidth]{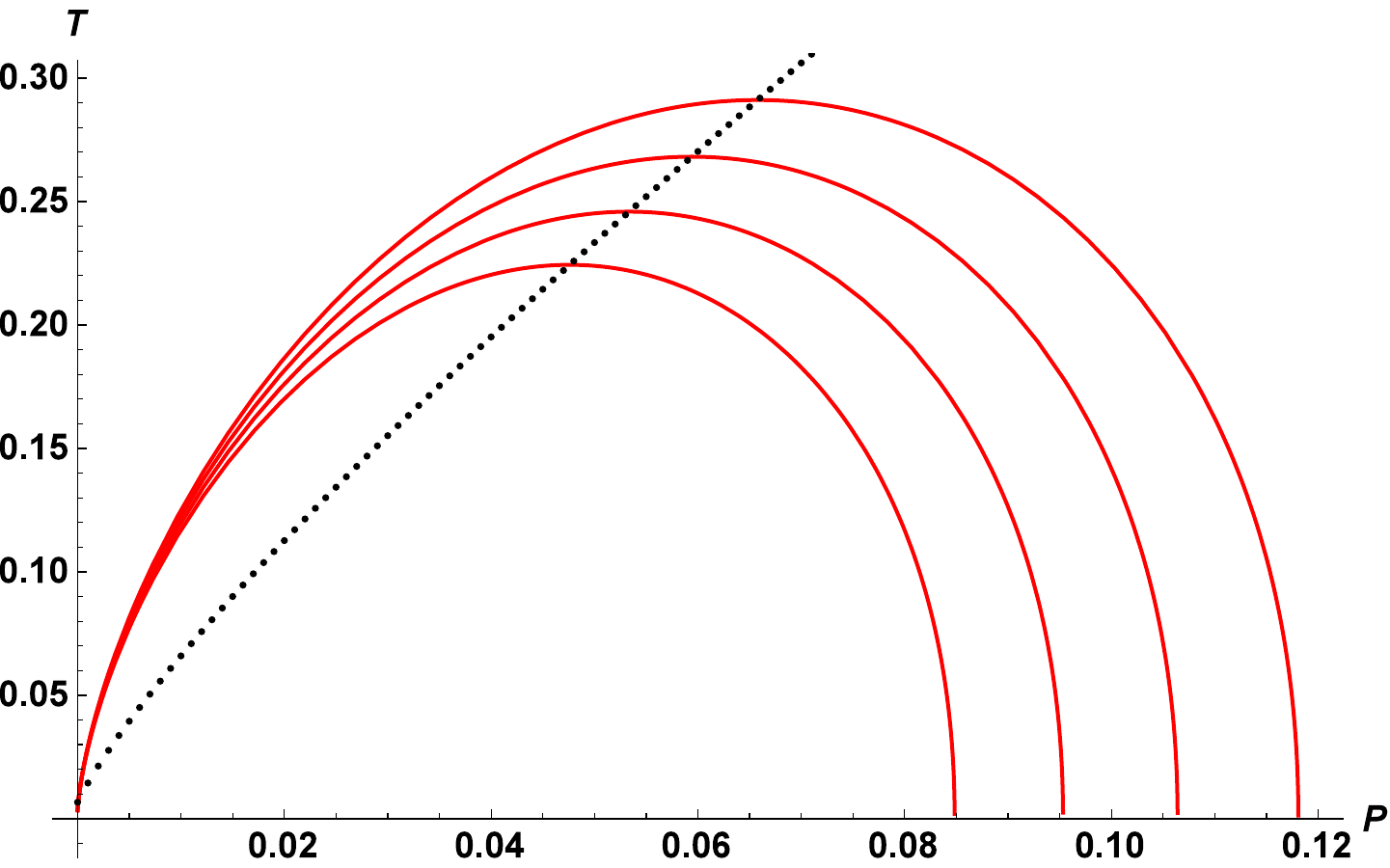}\hfil
	\includegraphics[width=0.47\linewidth]{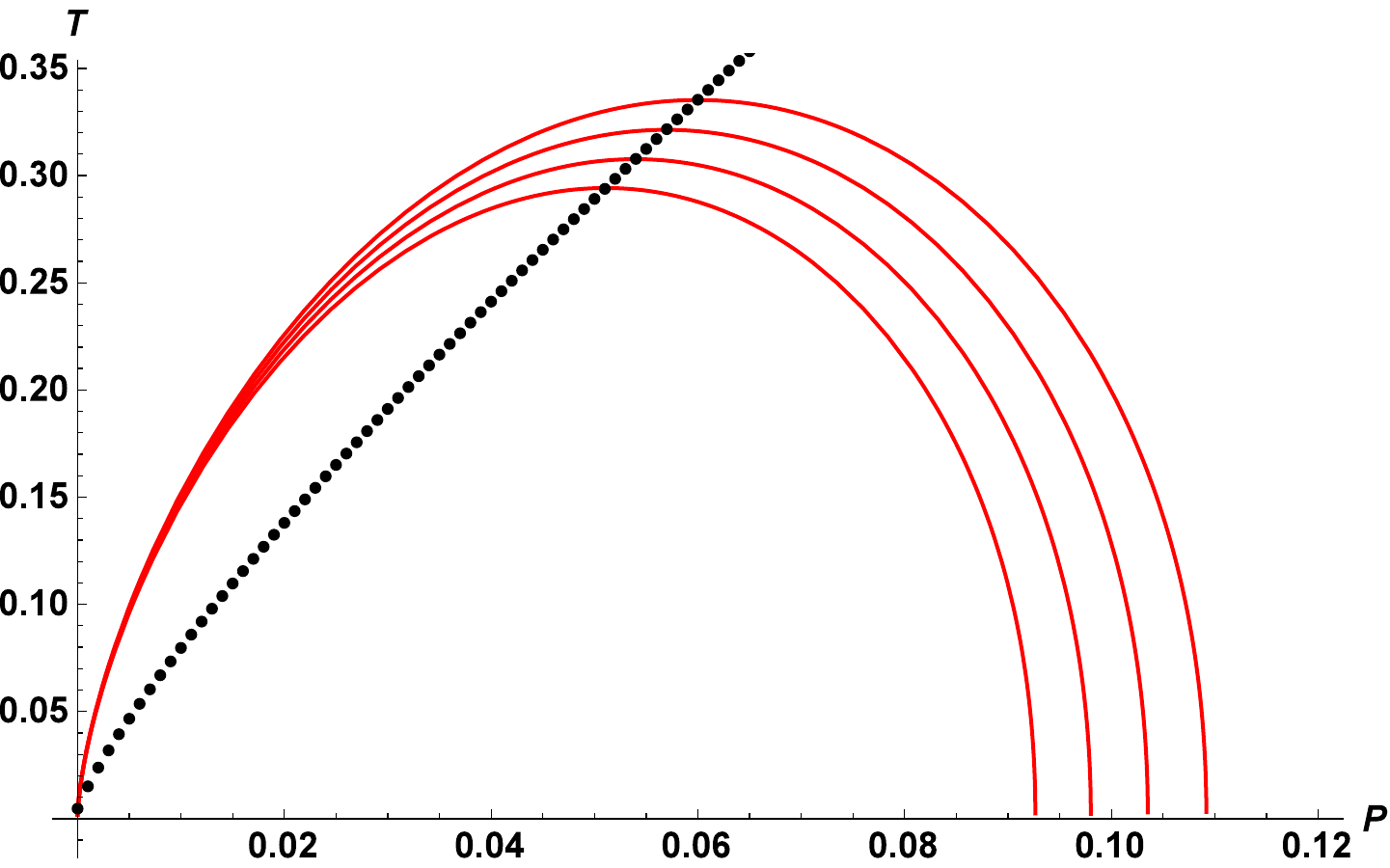}
	\caption{Inversion and isenthalpic (constant mass) curves of Kerr$-$AdS black holes. Red and black lines present isenthalpic and inversion curves, respectively. From bottom to top, isenthalpic curves correspond to increasing values of $M$. (Top-left) $J=1$ and $M=1.5,2.2.5,3$. (Top-right) $J=2$ and $M=2.5,3,3.5,4$. (Bottom-left) $J=10$ and $M=10.5,11,11.5,12$. (Bottom-right) $J=20$ and $M=20.5,21,21.5,22$.}
	\label{isC}
\end{figure}
which is the same as the value of charged AdS black holes \cite{Okcu2017}. 

Solving Eq.(\ref{relationSP}) may not be analytically possible. Therefore, we use numerical solutions to plot inversion curves in $T-P$ plane. In Fig.(\ref{iC}), we plotted inversion curves for various angular momentum values. In contrast to van der Waals fluids, it can be seen from Fig.(\ref{iC}) the  inversion curves are not closed and there is only one inversion curve. We found the similar behaviours for the charged AdS black holes in our previous work \cite{Okcu2017}. 

In Fig.(\ref{isC}), we plot isenthalpic (constant mass) and inversion curves for various values of angular momentum in $T-P$ plane. If entropy from Eq.(\ref{mass}) can be substituted into  Eq.(\ref{Tempature}), we obtain constant mass curves in $T-P$ plane. As it can be seen from Fig.(\ref{isC}), the inversion curves divide the plane into two regions. The region above the inversion curves corresponds to cooling region, while the region under the inversion curves corresponds to heating region. Indeed, heating and cooling regions are already determined from the sign of isenthalpic curves slope. The sign of slope is positive in the cooling region and it changes in the heating region. On the other hand, cooling (heating) does not happen on the inversion curve which plays role as a boundary between two regions.
\begin{figure}[h]
	\centering
	\includegraphics[width=1\linewidth]{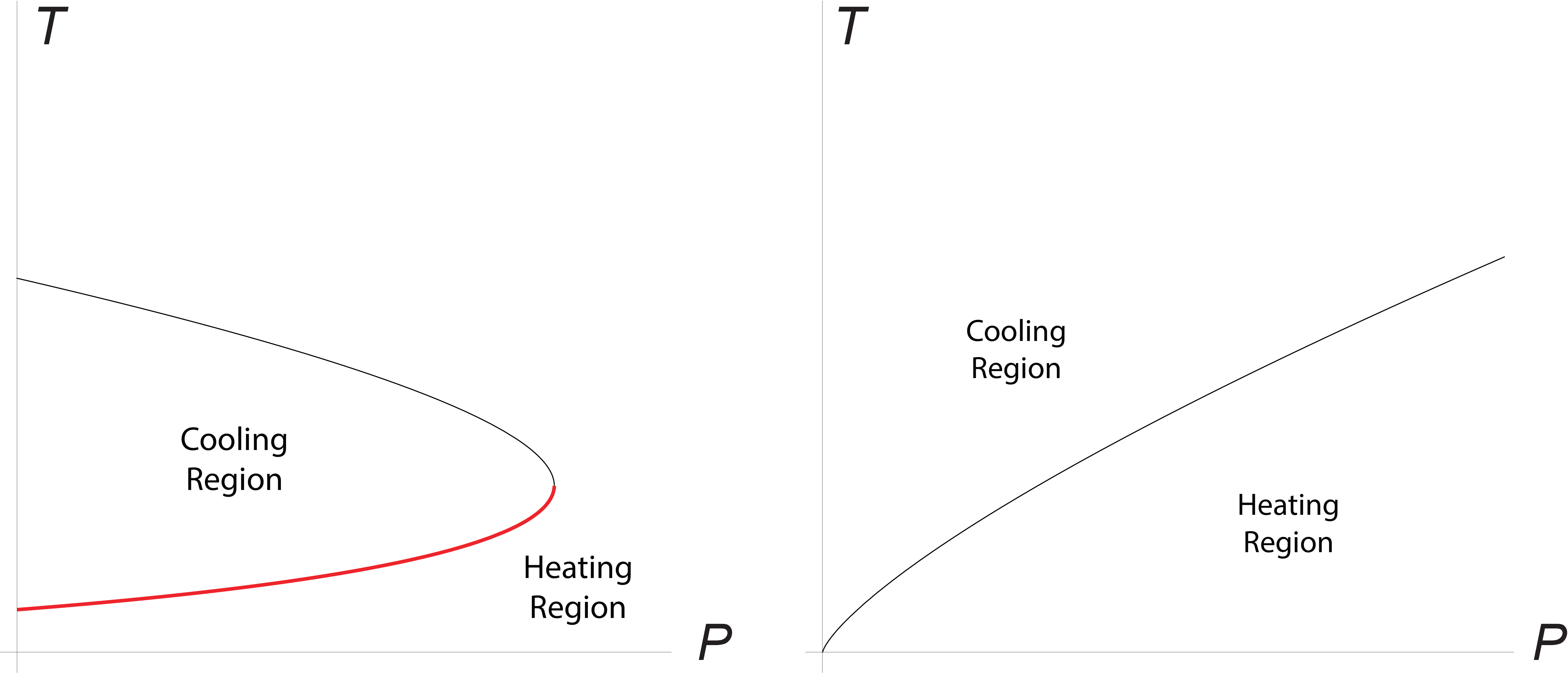}
	\caption{(Left) Inversion curves for van der Waals fluids. Red thick line and black solid line correspond to lower and upper inversion curves, respectively. (Right) Lower inversion curves for the charged AdS and Kerr-AdS black holes. }
	\label{iC2}
\end{figure}

\section{Conclusions}
In this study, we investigated Joule$-$Thomson expansion for Kerr$-$AdS black holes in the extended phase space. The Kerr-AdS black hole Joule$-$Thomson formula was derived by using the first law of black hole thermodynamics and Smarr relation. We plotted isenthalpic and inversion curves in $T-P$ plane. In order to plot inversion curves, we solved Eq.(\ref{relationSP}) numerically. Moreover, we obtained minimum inversion temperature $T_{i}$ and calculated ratio between inversion and critical temperatures for Kerr$-$AdS black holes.

Similar results were reported for the charged AdS black holes in \cite{Okcu2017} by us. For example, there is only a lower inversion curve  for Kerr$-$AdS and the charged AdS black holes. Therefore, we only consider minimum inversion temperature $T_{i}^{min}$ at $P_{i}=0$. Cooling regions are not closed for both systems. The ratios between minimum inversion temperatures and critical temperatures are nearly the same for the both black hole solutions. The ratio may deviates from $0.5$ for other black hole solutions. The same ratio may be obtained for other black hole solutions in the different limit cases. Furthermore, we restricted the study to four dimensional solution. Therefore the ratio may depend on dimensions  of space$-$time.

In order to compare the charged AdS$/$Kerr-AdS black holes with van der Waals fluids, we present schematic inversion curves for van der Waals fluids and the charged AdS/Ker-AdS black holes in Fig.(\ref{iC2}). In contrast to the charged AdS and Kerr-AdS black holes, there are upper and lower inversion curves for van der Waals fluids \cite{Okcu2017}. Therefore the cooling region is closed and we only consider both the minimum inversion temperature $T_{i}^{min}$ and maximum inversion temperature $T_{i}^{max}$ for this system. While cooling always occurs above the inversion curves for both black hole solutions, cooling only occurs in the region surrounded by the upper and lower inversions curves for van der Waals fluids.
\label{Con}
\acknowledgments
	We would like to thank the anonymous referees for their helpful and constructive comments.




\end{document}